\documentclass[12 pt, letter]{article}
\usepackage{graphicx}
\usepackage{caption}
\usepackage{amsmath}
\usepackage{amsthm}
\usepackage{enumerate}
\usepackage{subfig}
\usepackage{float}
\usepackage{amsfonts}
\usepackage{amssymb}
\usepackage{anysize}
\usepackage{helvet}
\usepackage{setspace}
\usepackage{multirow}
\usepackage{tabularx}
\usepackage{tabulary}
\usepackage{longtable}
\usepackage{float}
\usepackage{mathtools}
\title{\bf Ambiguity appearing on the Hamiltonian formulation of Quantum Mechanics.}
\author{Gustavo V. L\'opez\footnote{gulopez@cencar.udg.mx}~~, Ana Griselda, and Carlos R. Mart\'{\i}nez-Prieto\\ \\
 Departamento de F\'{i}sica, Universidad de Guadalajara,\\
 Blvd. Marcelino Garc\'{i}a Barragan y Calzada Ol\'{i}mpica, \\Ê44200 Guadalajara, Jalisco, Mexico}
\begin{document}
\maketitle

%\centerline{\large\bf Abstract}
\begin{abstract}
\noindent
For the symmetric harmonic oscillator and the symmetric bouncer defined in 2-D, two different Hamiltonian are given 
describing the same classical dynamics; however, their quantum dynamics behavior are different.
\end{abstract}
\vskip3pc
{\bf PACS :} 03.65.Ca, 03.65.Ta
\newpage
\section{ Introduction}
Modern Physics is based on Lagrangian or Hamiltonian [1,2,3] as mathematical objects to formulate and study the correspondent 
behavior of the natural systems. In particular, Quantum Mechanics has its foundations based on the Hamilton operator (generated from the 
Hamilton function of Classical Mechanics) associated to the studying system, and the Schr\"odinger's equation which describes the linear
evolution of the wave function. For most of the systems (conservative or including electromagnetic interaction) seems to be that there is not 
problem to get a well unique Hamiltonian formulation [4]. However, when dealing with dissipative systems, it appears some problems [5,6], 
and it has been shown [7,8] that even there can be two different Hamiltonian bringing about the same classical behavior but different quantum behavior. 
In this paper, we want to point out that this type of ambiguity also appears for two conservative symmetric systems with two degrees of freedom: the harmonic 
oscillator and the bouncer. Firstly, we make the deduction of the pair of Hamiltonian for both systems, each pair of them describing the same classical 
dynamics. Then, we show that their associated quantum dynamics are different for each system. 
\section{Classical Hamiltonians}
For the symmetric harmonic oscillator with two degrees of freedom, it is well known that a constant of motion and Hamiltonian can be given by
\begin{subequations}
\begin{equation}
K_1=\frac{m}{2}(v_x^2+v_y^2)+\frac{1}{2}m\omega^2(x^2+y^2)
\end{equation}
and
\begin{equation}\label{1h}
H_1=\frac{p_x^2+p_y^2}{2m}+\frac{1}{2}m\omega^2(x^2+y^2),
\end{equation}
\end{subequations}
where $m$ is the mass of the particle, and $\omega$ its angular frequency, being the same in both directions. However, it is not difficult to see that 
\begin{equation}\label{k2}
K_2=mv_xv_y+m\omega^2xy
\end{equation} 
is also a constant of motion ($dK_2/dt=0$) since it leads us to the equations of motion $m\dot{v}_x+m\omega^2 x=0$ and $m\dot{v}_y+m\omega^2y=0$. Once
we have this constant of motion, we know the known expression [9,10,11],
\begin{equation}\label{la}
L_2=v_x\int^{v_x}\frac{K_2(x,y,v_x,cv_x)}{v_x^2}dv_x,
\end{equation}
to get the Lagrangian knowing the constant of motion. By substituting (\ref{k2}) in (\ref{la}), one gets
\begin{equation}\label{l2}
L_2=mv_xv_y-m\omega^2xy,
\end{equation}
and the associated Hamiltonian ($H_2=v_xp_x+v_yp_y-L_2$) is
\begin{equation}\label{2h}
H_2=\frac{p_xp_y}{m}+m\omega^2xy.
\end{equation}
In this way, we have obtained two different Hamiltonians which describe the same classical dynamics.\\Ê\\
Consider now a symmetric bouncer with two degrees of freedom, characterized by the equations of motion
\begin{equation}\label{2s}
m\ddot x=-f,\quad\quad\hbox{and}\quad\quad m\ddot y=-f,
\end{equation} 
where $f$ is a constant force, and $m$ is the mass of the particle. The well known constant of motion and Hamiltonian are [13]
\begin{subequations}
\begin{equation}
\tilde K_1=\frac{m}{2}(v_x^2+v_y^2)+f(x+y)
\end{equation}
and
\begin{equation}\label{oneH}
\tilde H_1=\frac{p_x^2+p_y^2}{2m}+f(x+y).
\end{equation}
\end{subequations}
Again, it is not difficult to see that 
\begin{equation}
\tilde K_2=mv_xv_y+f(x+y)
\end{equation}
is also a constant of motion of the system (\ref{2s}). Then, using (\ref{la}), one gets the Lagrangian
\begin{equation}
\tilde L_2=mv_xv_y-f(x+y).
\end{equation}
The Hamiltonian of the system is now
\begin{equation}\label{twoH}
\tilde H_2=\frac{p_xp_y}{m}+f(x+y).
\end{equation}
So, once again we have obtained two different Hamiltonian which describe the same classical dynamics.
\section{Quantization}
The symmetric 2-D harmonic oscillator and symmetric 2-D bouncer represent autonomous systems (Hamiltonians do not depend explicitly on time). So, solving 
the Schr\"odinger's equation,
\begin{equation}
i\hbar\frac{\partial|\Psi\rangle}{\partial t}=\widehat H|\Psi\rangle,
\end{equation} 
for the associated Hermitian Hamiltonian operator $\widehat H$, is reduce to an eigenvalue problem
\begin{equation}
\widehat H|\Phi\rangle=E|\Phi\rangle,
\end{equation}
through the transformation
\begin{equation}
|\Psi\rangle=e^{-i Et/\hbar}|\Phi\rangle.
\end{equation}
The solution of the 2-D symmetric harmonic oscillator eigenvalue problem is well known, and it is given by
\begin{equation}\label{ev1}
E_{n_1n_2}^{(1)}=\hbar\omega(n+1),\quad\quad \langle{\bf x}|n_1n_2\rangle=\prod_{i=1}^2c_{n_i}e^{-\alpha x^2/2}H_{n_i}(\alpha x_i),
\end{equation}
where $n$ is a non negative  integer, $n=n_1+n_2\in{\cal Z}^+$, $c_{n_i}$ is a constant, $c_{n_i}=\sqrt{\alpha/\sqrt{\pi}~2^{n_i}n_i!}$, and $\alpha$ is the constant defined as $\alpha=\sqrt{m\omega/\hbar}$. The full solution is 
\begin{equation}\label{fs1}
|\Psi\rangle=\sum_{n_1,n_2=0}C_{n_1n_2}e^{-i\omega(n+1)t}|n_1n_2\rangle, \quad\quad\sum_{n_1,n_2=0}|C_{n_1n_2}|^2=1,
\end{equation}
where the coefficient $C_{n_1n_2}$ represents the amplitude of probability and can be determined by the initial conditions. 
The system is degenerated because of the symmetry, and the ground state is the state $|00\rangle$. \\Ê\\
For the Hamiltonian (\ref{2h}), let us make the following change of variables
\begin{equation}\label{newv}
\xi=\frac{1}{\sqrt{2}}(x+y), \quad\quad\hbox{and}\quad \eta=\frac{1}{\sqrt{2}}(x-y).
\end{equation}
So, the Hermitian Hamiltonian becomes
\begin{equation}\label{n2h}
\widehat H_2=\frac{\hat p_{\xi}^2-\hat p_{\eta}^2}{2m}+\frac{}{2}m\omega^2(\xi^2-\eta^2),
\end{equation}
where the operators $\hat p_{\xi}$ and $\hat p_{\eta}$ are defined as $\hat p_{\xi}=-i\hbar\partial/\partial\xi$ and $\hat p_{\eta}=-i\hbar\partial/\partial\eta$. This Hamiltonian is just the difference of two 1-D symmetric harmonic oscillators (ho), $\widehat H_2=\widehat H_{\xi}^{(ho)}-\widehat H_{\eta}^{(ho)}$. This means that one has the following solution for the eigenvalue problem
\begin{equation}\label{ev2}
E_{n_1n_2}^{(2)}=\hbar\omega(n_1-n_2),\quad\quad \langle{\bf x}|n_1n_2\rangle=\Phi_{n_1}\left(\frac{x+y}{\sqrt{2}}\right)\Phi_{n_2}\left(\frac{x-y}{\sqrt{2}}\right),
\end{equation}
where the function $\Phi_n(z)$ is defined as $\Phi_n(z)=c_{n}e^{-\alpha x^2/2}H_{n}(\alpha z)$, with the constants $c_n$ and $\alpha$ defined as before.Denoting the state defined by the product of the function of (\ref{ev2}) as $|n_1n_2\rangle_{_2}$, the full solution is written as
\begin{equation}\label{fs2}
|\Psi\rangle=\sum_{n_1,n_2=0}D_{n_1n_2}~e^{-i\omega(n_1-n_2) t}|n_1n_2\rangle_{_2}, \quad\quad\sum_{n_1,n_2=0}|D_{n_1n_2}|^2=1.
\end{equation}
The degeneration is infinity and there is not ground state of the system. Therefore, the quantum dynamics describe by (\ref{fs1}) and (\ref{fs2}) are completely different.\\Ê\\
For the usual Hamiltonian associated to the bouncer (\ref{oneH}), one sees that this one can be written of the form $\widetilde H_1=\widetilde H_x+\widetilde H_y$, where $\widetilde H_x=p_x^2/2m+fx$ and $\widetilde H_y=p_y^2/2m+fy$ correspond to the bouncer on each direction. The quantum bouncer eigenvalues are given in terms of the zeros of the Airy function [13]. Therefore, the solution of the eigenvalue problem $\widehat{\widetilde H}_1|\Phi\rangle=E|\Phi\rangle$ is given by
\begin{equation}
\widetilde{E}_{n_1n_2}^{(1)}=fl_f(z_{n_1}+z_{n_2}),\quad\quad Ai(-z_{n_i})=0, \quad\quad l_f=\left(\frac{\hbar^2}{2mf}\right)^{1/3}.
\end{equation}  
The eigenfunctions are
\begin{equation}
\widetilde{\Phi}_{n_1n_2}=\langle{\bf x}|n_1n_2\rangle=\frac{Ai(z_1-z_{n_1})}{|A'i(-z_{n_1})|}\cdot\frac{Ai(z_2-z_{n_2})}{|A'i(-z_{n_2})|},
\end{equation}
where $z_1$ and $z_2$ are defined as $z_1=x/l_f$ and $z_2=y/l_f$, $A'i$ represents the differentiation of the Airy function. The full Scr\"odinger solution is
\begin{equation}\label{endb1}
|\Psi\rangle=\sum_{n_1,n_2}\widetilde C_{n_1n_2}e^{-ifl_f(z_{n_1}+z_{n_2})t/\hbar}~|n_1n_2\rangle, \quad\quad \sum_{n_1n_2}|C_{n_1n_2}|^2=1.
\end{equation}
The spectrum is discrete,  there is degeneration (because of the symmetry), and there is a ground state for the system.\\Ê\\
Now, for the bouncer Hamiltonian (\ref{twoH}), we make the same change of variables (\ref{newv}) which brings about the Hamiltonian operator
\begin{equation}
\widehat H_2=\frac{\hat p_{\xi}^2-\hat p_{\eta}^2}{2m}+f\sqrt{2}~\xi,
\end{equation}
corresponding a quantum bouncer in the variable $\xi$, and free particle quantum motion in the variable $\eta$. The eigenvalues of this Hamiltonian operators are
\begin{equation}
\widetilde E_{nk}^{(2)}=\sqrt{2}fl_f^*z_n+\frac{\hbar^2 k^2}{2m},\quad\quad l_f^*=\left(\frac{\hbar^2}{2\sqrt{2}~mf}\right)^{1/3},
\end{equation}
where $z_n$ is the zero of the Airy function, and $k$ is a continuous real constant. The eigenfunctions are
\begin{equation}
\widetilde \Phi_{nk}=\langle{\bf x}|nk\rangle=\frac{Ai(z-z_{n})}{|A'i(-z_{n})|}~e^{ik(x-y)/\sqrt{2}},
\end{equation} 
where the variable $z$ is $z=\xi/l_f^*=(x+y)/l_f^*\sqrt{2}$, and the full solution of the Schr\"odinger's equations is
\begin{equation}\label{endb2}
|\Psi\rangle=\sum_n\int dk C_n(k)e^{-i\widetilde{E}_{nk}^{(2)}}~|nk\rangle, \quad\quad\quad\quad \sum_n\int dk |C_n(k)|^2=1.
\end{equation}
The spectrum has a discrete component and a continuous component, and there is a ground state of the system for $k=0$ and $n=1$. As one can see from (\ref{endb1}) and (\ref{endb2}) the quantum dynamics is different.
\section{Conclusion}
We have shown that for two conservative 2-D symmetrical systems (harmonic oscillator and bouncer), we can find for each of them at least two different Hamiltonians describing the same classical dynamics but different quantum dynamics. It is our guess that this type of ambiguity is intrinsic of the Hamiltonian 
theory, and it could be present on any quantum system. 
\newpage\indent
{\bf \Large References}\\Ê\\
1. C. Cohen-Tannoudji, B. Diu, and F. Lalo\"e, {\it Quantum Mechanics,I, II,} John Wiley\&Sons, (1977).\\Ê\\
2. H. Huang, {\it Statistical Mechanics,} John Wiley N.Y. (1963).\\Ê\\
3. S.S. Schweber, {\it An Introduction to Relativistic Quantum Field Theory,} Dover Publications Inc., (2005).\\Ê\\
4. P.A.M. Dirac, {\it The Principles of Quantum Mechanics,} Oxford University Press, (11976).\\Ê\\
5. R. Glauber and V.I.Man'ko, Soviet Physics JETP, {\bf 60}, (1984),450.\\Ê\\
6. G. L\'opez, Int. J. Theo. Phys.,{\bf 37}, (1998), 1617. \\Ê\\
7. V.P. Dodonov, V.I. Man'ko and V.D. Skarzhinsky, Hadronic Journal, {\bf 4}, (1981),1734.\\Ê\\
8. G. L\'opez, P. L\'opez, and X.E. L\'opez, Adv. Studies Theor. Phys., {\bf 5}, (2011), 253.\\Ê\\
9. J.A. Kobusen, Act. Phys. Austr.,{\bf 51}, (1979), 193.\\Ê\\
10. C. Leubner, Phys. Rev. A, {\bf 86}, (1987), 9.\\Ê\\
11. G. L\'opez, Ann, Phys., {\bf 251}, (1996), 372.\\Ê\\
12. A. Messiah, {\it Quantum Mechanics, I, II}, North-Holland Publishing, (1958).\\Ê\\
13. G. L\'opez and G. Gonz\'alez, Int. J. Theo. Phys.,{\bf 43}, (2004), 1999.

-

\end{document}